**Overcoming noise in quantum teleportation with multipartite hybrid entanglement**

Zhao-Di Liu[1, 2, †], Olli Siltanen[3, 4, †], Tom Kuusela[3], Rui-Heng Miao[1, 2], Chen-Xi Ning[1, 2], Chuan-Feng Li[1, 2, 5, *, ‡], Guang-Can Guo[1, 2, 5], and Jyrki Piilo[3, *, §]

[1]CAS Key Laboratory of Quantum Information, University of Science and Technology of China, Hefei, 230026, China.

[2]CAS Center For Excellence in Quantum Information and Quantum Physics, University of Science and Technology of China, Hefei, 230026, China.

[3]Department of Physics and Astronomy, University of Turku, FI-20014 Turun yliopisto, Finland.

[4]Department of Mechanical and Materials Engineering, University of Turku, FI-20014 Turun yliopisto, Finland.

[5]Hefei National Laboratory, University of Science and Technology of China, Hefei, 230088, China.

[†]These authors contributed equally to this work.

[‡]cfli@ustc.edu.cn
[§]jyrki.piilo@utu.fi

**ABSTRACT**

Quantum entanglement and decoherence are the two counterforces of many quantum technologies and protocols. For example, while quantum teleportation is fueled by a pair of maximally entangled resource qubits, it is vulnerable to decoherence. In this Article, we propose an efficient quantum teleportation protocol in the presence of pure decoherence and without entangled resource qubits entering the Bell-state measurement. Instead, we employ multipartite hybrid entanglement between the auxiliary qubits and their local environments within the open-quantum-system context. Interestingly, with a hybrid-entangled initial state, it is the decoherence that allows us to achieve high fidelities. We demonstrate our protocol in an all-optical experiment.

1. **INTRODUCTION**

Quantum entanglement manifests itself in correlations that span over arbitrarily long distances (*1*). Besides its great significance on the foundations of quantum mechanics, entanglement has found numerous applications in quantum information processing and quantum communication, e.g., quantum teleportation (*2–7*), superdense coding (*8–10*), and quantum key distribution (*11–13*). However, the unavoidable interactions between a quantum system and its environment can severely degrade the performance of these applications by means of decoherence (*14*). Avoiding any kind of decoherence is extremely demanding in practice, even though many promising decoherence suppression protocols have been proposed. Some recent works have exploited delayed coherent quantum feedback (*15–17*), reservoir engineering with auxiliary subsystems (*18–20*), quantum error-correcting codes (*21–23*), dynamical decoupling (*24–26*), and decoherence-free subspaces (*27, 28*).

Linear optics provides a particularly robust platform in which to perform different quantum information protocols and study the problematics with decoherence. Here, the system of interest often consists of the polarization degrees of freedom of individual photons, their frequency represents the environment, and the system-environment interaction is realized controllably in birefringent media (*3, 7, 9, 27–32*). Although the total dynamics is unitary, the open system undergoes nonunitary dynamics, which is obtained by averaging over the environment. The resulting dephasing drives the coherence terms of a given system to zero while keeping its populations intact, corresponding to quantum-to-classical transition.

In this Article, we study quantum teleportation under dephasing in the afore described linear optical framework. We attack dephasing by employing multipartite hybrid entanglement. Hybrid entanglement means entanglement between different degrees of freedom (*33, 34*), and it has been previously utilized in overcoming the probabilistic nature of discrete-variable teleportation (*35*) but not yet in our context. With hybrid entanglement, we can controllably scramble the open system's phase information so that the subsequent dephasing later reassembles it instead of scrambling it even more. As a consequence, with dephasing appearing in the end of our teleportation protocol, we can transform system-environment correlations into coherences within the teleported state. Note that throughout this Article, we will use "decoherence" and "dephasing" irrespective of how the coherences evolve.

Remarkably, we achieve high fidelities without the resource qubits being entangled in the Bell-state measurement. Moreover, our protocol works without any frequency correlations. Although the environment in our case is controllable, our proof-of-concept work lays the groundwork for future technologies with the anticipation that as technology advances, similar control may be exerted over more realistic environmental factors.

## 2. RESULTS

**Theoretical Description:**

The standard quantum teleportation protocol goes as follows (*2*). Alice has a qubit whose unknown state $|\phi\rangle = \alpha|0\rangle + \beta|1\rangle$ she wants to teleport to Bob. In view of this task, Alice and Bob have shared a Bell state. Alice performs a Bell-state measurement (BSM) on her

pair of states, i.e., the one to be teleported and her part of the auxiliary state. The two become fully entangled, erasing the initial entanglement between Alice and Bob due to the monogamy of entanglement. Alice classically reports her result to Bob, who then applies a unitary on his state, matching with the reported Bell state. As a result, Bob's state becomes $|\phi\rangle$.

The standard protocol assumes ideal conditions, i.e., no noise. We now consider a more realistic scenario, where the auxiliary state shared by Alice and Bob experiences local dephasing noise for the respective durations of $T_a$ and $T_b$. The steps of our protocol are illustrated in Fig. 1.

Encoding the auxiliary qubits into the polarization degrees of freedom of two photons, having their frequencies act as local environments, and taking initial system-environment correlations into account, the total state of the two photons preceding the noise reads

$$|\Psi(0,0)\rangle = \frac{1}{\sqrt{2}}\{|HV\rangle \int df_a df_b g(f_a, f_b) e^{i[\theta_{aH}(f_a) + \theta_{bV}(f_b)]}$$
$$+ |VH\rangle \int df_a df_b g(f_a, f_b) e^{i[\theta_{aV}(f_a) + \theta_{bH}(f_b)]}\}|f_a f_b\rangle. \qquad (1)$$

Here, H (V) denotes horizontal (vertical) polarization, $g(f_a, f_b)$ is the probability amplitude corresponding to the joint frequency state $|f_a f_b\rangle$ satisfying the normalization condition $\int df_a df_b |g(f_a, f_b)|^2 = 1$, and the $\theta$-functions describe the initial polarization-frequency correlations. Having access to these functions Alice and Bob can tailor the correlations to their liking. We make no other assumptions, e.g., about the frequency distribution.

The decoherence that the two polarization qubits will experience is described by the system-environment interaction Hamiltonian $\mathbb{H}_a \otimes \mathbb{I}_b + \mathbb{I}_a \otimes \mathbb{H}_b$, where the local Hamiltonians are of the pure-dephasing form (*14*)

$$\mathbb{H}_j = -(n_{jH}|H\rangle\langle H| + n_{jV}|V\rangle\langle V|) \otimes \int df_j 2\pi f_j |f_j\rangle\langle f_j|. \qquad (2)$$

Here, $n_{j\lambda}$ is the refractive index of a birefringent medium corresponding to Alice ($j = a$) or Bob's ($j = b$) polarization component $\lambda = H, V$. This leads to the joint unitary evolution of the system and environment, $U_j(t_j)|\lambda\rangle|f_j\rangle = e^{i2\pi f_j n_{j\lambda} t_j}|\lambda\rangle|f_j\rangle$, with $t_j$ denoting the interaction time. Because the open system and its environment together form a closed system, no information is truly lost regardless of initial correlations. This comes into play later. To take this kind of noise into account, Alice and Bob fix

$$\theta_j(f_j) = \theta_{jH}(f_j) - \theta_{jV}(f_j) = -2\pi f_j \Delta n_j T_j \qquad (3)$$

independently of each other. Here, $\Delta n_j = n_{jH} - n_{jV}$ denotes the birefringence. Physically, imposing condition (3) on the state (1) entangles its (composite) polarization with the (composite) frequency—creating a multipartite hybrid-entangled state (*33–35*)— and this can be done, e.g., with spatial light modulators [SLMs (*32*)]. How this hybrid entanglement is distributed between Alice and Bob depends on the initial probability amplitudes. With our choices, for example, there is no entanglement in the open system or its environment alone, but rather, the local system-environment states of Alice and Bob are entangled with each other (see the Supplementary Materials for more details).

Before having any noise, the bipartite polarization state, obtained by taking partial trace of $|\Psi(0,0)\rangle\langle\Psi(0,0)|$ over frequency, reads

$$\rho_{ab}(0,0) = \frac{1}{2}\begin{pmatrix} 0 & 0 & 0 & 0 \\ 0 & 1 & \Lambda_{ab}(0,0) & 0 \\ 0 & \Lambda_{ab}^*(0,0) & 1 & 0 \\ 0 & 0 & 0 & 0 \end{pmatrix}, \quad (4)$$

where $\Lambda_{ab}(t_a, t_b) = \int df_a df_b |g(f_a, f_b)|^2 \exp\{i[\theta_a(f_a) + 2\pi f_a \Delta n_a t_a - \theta_b(f_b) - 2\pi f_b \Delta n_b t_b]\}$ is the bivariate decoherence function governing the decoherence dynamics of the auxiliary state. With $T_j \gg 0$, we obtain the approximate form $\rho_{ab}(0,0) \approx (|HV\rangle\langle HV| + |VH\rangle\langle VH|)/2$ that is (seemingly) local by its nature, i.e., its nonlocality is "hidden".

Alice then removes her contribution of the polarization-frequency correlations described by $\theta_a(f_a) = -2\pi f_a \Delta n_a T_a$ by letting her auxiliary photon go through a birefringent crystal with the birefringence $\Delta n_a$ and length $cT_a$. It should be stressed that it need not be Alice who operates the SLM and implements noise. It might as well be a third party that prepares the auxiliary photons and sends one to Alice via noisy channel.

Because $T_b \gg 0$, we still have the decoherence function $\Lambda(T_a, 0) \approx 0$ and therefore the mixed state $\rho_{ab}(T_a, 0) \approx (|HV\rangle\langle HV| + |VH\rangle\langle VH|)/2$. On the other hand, denoting the state being teleported by $|\phi\rangle = \alpha|H\rangle + \beta|V\rangle$, the total state of the *three* photons is

$$|\Omega(T_a, 0)\rangle = \frac{1}{\sqrt{2}}|\phi\rangle[|HV\rangle|\xi_{HV}(T_a, 0)\rangle + |VH\rangle|\xi_{VH}(T_a, 0)\rangle], \quad (5)$$

where $|\xi_{\lambda\lambda'}(T_a, 0)\rangle = \int df_a df_b g(f_a, f_b) \exp\{i[\theta_{a\lambda}(f_a) + 2\pi f_a n_{a\lambda} T_a + \theta_{b\lambda'}(f_b)]\}|f_a f_b\rangle$. Note that, in Eq. (5), we have written the state to be teleported first, then Alice's auxiliary qubit, and finally Bob's auxiliary qubit, and that we will keep this order in the following. Namely, Eq. (5) can also be written in the form

$$\begin{aligned}|\Omega(T_a, 0)\rangle = &\frac{1}{2}|\Phi^+\rangle[\beta|H\rangle|\xi_{VH}(T_a, 0)\rangle + \alpha|V\rangle|\xi_{HV}(T_a, 0)\rangle] \\ &+ \frac{1}{2}|\Phi^-\rangle[-\beta|H\rangle|\xi_{VH}(T_a, 0)\rangle + \alpha|V\rangle|\xi_{HV}(T_a, 0)\rangle] \\ &+ \frac{1}{2}|\Psi^+\rangle[\alpha|H\rangle|\xi_{VH}(T_a, 0)\rangle + \beta|V\rangle|\xi_{HV}(T_a, 0)\rangle] \\ &+ \frac{1}{2}|\Psi^-\rangle[\alpha|H\rangle|\xi_{VH}(T_a, 0)\rangle - \beta|V\rangle|\xi_{HV}(T_a, 0)\rangle], \quad (6)\end{aligned}$$

where $|\Phi^\pm\rangle = (|HH\rangle \pm |VV\rangle)/\sqrt{2}$ and $|\Psi^\pm\rangle = (|HV\rangle \pm |VH\rangle)/\sqrt{2}$ are the Bell states. Alice performs BSM on her pair of qubits. This simultaneously entangles her polarization qubits and, more importantly, remotely transforms the hybrid entanglement into local system-environment entanglement on Bob's side. Alice communicates her result $|B\rangle$ to Bob, who then applies a matching unitary operation $U_B$ on his qubit $\rho_b(0)$. The matching unitaries are

$$U_B = \begin{cases} \sigma_x \text{ for } |\Phi^+\rangle, \\ i\sigma_y \text{ for } |\Phi^-\rangle, \\ \mathbb{1} \text{ for } |\Psi^+\rangle, \\ \sigma_z \text{ for } |\Psi^-\rangle. \end{cases} \tag{7}$$

Now, Bob's polarization state reads

$$U_B \rho_b(0) U_B^\dagger = \begin{pmatrix} |\alpha|^2 & \alpha\beta^* \Lambda_{ab}(T_a, 0) \\ \alpha^*\beta \Lambda_{ab}^*(T_a, 0) & |\beta|^2 \end{pmatrix}. \tag{8}$$

It is important to note that in the above equation, Bob's qubit's decoherence function is now exactly the one that previously described the nonlocal coherences for the auxiliary qubit pair. It still has the value $\Lambda_{ab}(T_a, 0) \approx 0$ [see Eq. (4) and text below that]—while Alice's BSM has remotely entangled Bob's polarization and frequency. Finally, Bob can retrieve the previously hidden information and purify the state (8) by subjecting it to dephasing noise determined by the effective path difference $c\Delta n_b T_b$. With $\Lambda_{ab}(T_a, T_b) = 1$, the final state becomes $|\phi\rangle$. Note that Bob needs no information about Alice's system-environment correlations.

Remarkably, the biphoton polarization shared by Alice and Bob does not violate the Bell-CHSH inequalities during BSM (for details, see the Supplementary Materials). Furthermore, our protocol works even with uncorrelated frequencies, i.e., with $g(f_a, f_b) = g_a(f_a) g_b(f_b)$, as opposed to prior teleportation schemes using active nonlocality and fully anticorrelated frequencies to battle decoherence [see, e.g., Ref. (*7*)]. Hence, our protocol is not limited to using specific initial polarization states. We did not assume anything about the paths leading to Alice and Bob either; Accounting for free evolution, the two time variables in the decoherence function $\Lambda(t_a, t_b)$ would simply become $t_j \mapsto t_j + \Delta n_{\text{air}}/\Delta n_j t_{j,\text{free}}$, but because $\Delta n_{\text{air}} \approx 0$, we immediately see that free evolution does not influence our protocol.

Recently, a teleportation protocol starting from a classical two-qubit state was demonstrated in (*36*), where the "hidden nonlocality" of the two auxiliary qubits was first activated with local filters. Our protocol too fits well with the concept of hidden nonlocality (*37–39*), albeit our filters are unitaries. Yet more importantly, our protocol shows that nonlocality *need not be fully activated* before the BSM. As a consequence, the nonlocality of the polarization remains always hidden.

**Experimental Results:**

The three photons needed in the protocol were prepared in two consecutive spontaneous parametric down-conversion (SPDC) processes. The phase functions were implemented with SLMs and noise with birefringent crystals. Finally, the BSM was carried out in standard manner with linear optical elements. The experimental setup is presented in Fig. 2, and further details are given in Experimental Design, under Materials and Methods.

To thoroughly test our protocol, we teleported the states $|+\rangle = (|H\rangle + |V\rangle)/\sqrt{2}$, $|-\rangle = (|H\rangle - |V\rangle)/\sqrt{2}$, $|R\rangle = (|H\rangle + i|V\rangle)/\sqrt{2}$, and $|L\rangle = (|H\rangle - i|V\rangle)/\sqrt{2}$ using different versions of it. In each case, noise was implemented either on Alice's side ($400\lambda_0$ of YVO4), Bob's side ($411\lambda_0$ of quartz), or both. In addition, we either used SLMs (i.e., hybrid entanglement) or not. Whenever SLMs were used, Alice's phase function was $\theta_a(f_a) = -2\pi f_a/c \cdot 446\lambda_0$ and Bob's $\theta_b(f_b) = -2\pi f_b/c \cdot 429\lambda_0$. The factors in front of $\lambda_0$ in the phase functions were carefully optimized to mitigate dispersion in the birefringent crystals. This explains the mismatch between the amount of noise and the said factors (400/446 and 411/429).

Fig. 3 shows the fidelities of the final states $\rho_f$ with respect to the input states $\rho_i$, given by $F(\rho_f, \rho_i) = \left(\text{tr}\sqrt{\sqrt{\rho_f}\rho_i\sqrt{\rho_f}}\right)^2$. The bars labeled by "A", "B", and "A+B" correspond to Alice's noise configuration, Bob's noise configuration, and the combination of these two, respectively. The lower, red bars give fidelities of the protocols *without* SLMs, whereas the higher, green bars give fidelities of the protocols *with* SLMs. The orange lines correspond to situations, where we had neither noise nor SLMs. The black dotted line is the classical average fidelity limit, 2/3 (*40*).

From Fig. 3, we can clearly see the problem with decoherence even with entangled polarization qubits (lower, red bars) and how hybrid entanglement helps us resolve it (higher, green bars). The green bars labeled by "A+B" and "B" describe the main results of this paper and correspond to the protocol starting from Figs. 1A and 1B, respectively. Here, the total state of the two photons prior to BSM is hybrid-entangled; From the open system's point of view, the nonlocality is hidden. Still, we achieve very high fidelities in the end, in all cases well above the classical average fidelity limit and approximately equal to the reference fidelities.

The green bars labeled by "A" correspond to standard teleportation after Alice's dephasing. Namely, Alice's dephasing converts the hybrid entanglement into typical polarization entanglement, giving Alice and Bob just the Bell state $|\Psi^+\rangle$ prior to BSM. Similar results are reported in the Supplementary Materials Section S2. There, we purified all the Bell states with SLMs and subsequent dephasing. In addition to YVO4 and quartz, we used a 2 m polarization-maintaining single mode fiber, corresponding better to a real-life scenario.

## 3. DISCUSSION

In this work, we have demonstrated noisy quantum teleportation with the help of multipartite hybrid entanglement. Against the traditional viewpoint of decoherence always acting as a drawback, here dephasing *helps* us achieve high-fidelity final states. This is due to hybrid entanglement effectively reversing the direction of decoherence. It allows us to start from a classical polarization state and end up with the desired quantum state. Consequently, the resource qubits need not violate the Bell-CHSH inequalities in BSM.

We emphasize that only the initial phase functions need to be controlled. Because neither the frequency distribution nor the frequency correlations play a role in our protocol, we can use any initial Bell state (without coherences) as the auxiliary qubits.

Hybrid entanglement not only helps with fighting decoherence. It can also bring another layer of security. Consider a general setting and Eve the Eavesdropper capturing Bob's qubit before Bob has purified it with dephasing. Even if Eve knew Alice's BSM result, she could not purify the captured qubit, because it is not correlated with Eve's environment. With dephasing, Eve could only make things worse. It would be an interesting line of future research to investigate how deep the teleported information can be hidden, i.e., how large hybrid-entangled total state we can use.

The experimental results presented in the Supplementary Materials suggest that our technique could also be applied in state transfer outside quantum teleportation. In theory, we could transfer *any N*-qubit state across dephasing environments, with a pure state exiting the network. Hence, we could go well beyond decoherence-free subspaces. In practice, only the resolution of the SLMs might limit such applications.

The main assumptions in this work were prior knowledge on the duration of dephasing and access to initial system-environment correlations. Not knowing the lengths of the dephasing channels, Alice and Bob would face an interesting optimization problem in terms of their phase functions. In fact, Fig. S4 corresponds to such a situation. Should the second assumption fail, it would be worth investigating if ancillary systems could help, e.g., like Greenberger-Horne-Zeilinger (GHZ) states in (*41, 42*).

In general, our results highlight the importance of state preparation in the applications of quantum theory and shed new light on entanglement recycling. While our work has a proof-of-principle character, it also opens the possibility to see if decoherence can be reversed in other physical platforms, including different sources of noise.

**MATERIALS AND METHODS**

**Experimental Design:**

First, a combination of beta barium borate (BBO) nonlinear crystal, half-wave plate (HWP), and another BBO (together "C-BBO" in Fig. 2) is pumped by a femtosecond ultraviolet laser (390 nm, 76 MHz). The entangled photon pairs produced by SPDC are distributed to the sides of Alice and Bob.

The photons that did not get down-converted proceed to one more BBO crystal. Here, the state to be teleported is created together with a photon that later triggers the coincidence counting electronics. The state being teleported is finalized by HWP2 and a quarter-wave plate (QWP2).

Next, in case any noise is to be implemented later, Alice and Bob's phase functions need to be imprinted on their auxiliary photons. This is achieved by guiding the photons through SLMs that are sandwiched between gratings (1200 l/mm), plano-convex cylindrical lenses (PCC lenses), beam displacers (BDs), and HWPs. The gratings and PCC lenses convert the frequency of the photons to spatial degrees of freedom. Because the SLMs are only effective for one polarization component (here for H), the BDs and 45° HWPs are used to change the polarization state into a path state (with H polarization). 150 SLM-pixels cover about 3.5 nm of the photons' spectra, i.e., their full width at half maximum (FWHM), so that the SLMs can accurately manipulate the phase functions at pixel level. For other pixels, we designed a phase function similar to blazed grating. Its purpose is to diffract excess photons to other angles. The center of the photon spectrum $\lambda_0$ is carefully aligned to the middle of the SLMs.

If the SLMs were used, the auxiliary photons are now in a multipartite hybrid-entangled state. After state preparation, Alice's photon is subjected to polarization dephasing in a YVO$_4$ crystal so that Alice's phase function and noise terms cancel each other. If the SLM was used on Bob's side, the total state is still hybrid-entangled.

Next, Alice performs BSM on her part of the auxiliary pair and the state being teleported. The BSM entangles Alice's polarization qubits and remotely transforms the multipartite hybrid entanglement shared by Alice and Bob to local system-environment entanglement on Bob's side. The BSM is achieved by HWP1, HWP3, HWP4, and three polarizing beam splitters (PBSs). When we measure the Bell states $|\Phi^\pm\rangle$, HWP1 is set to 0°, and HWP3 and HWP4 are set to ±22.5°. For $|\Psi^\pm\rangle$, HWP1 is set to 45°, and HWP3 and HWP4 are set to ±22.5°. In order to make the photonic identity better, we use 2-nm-FWHM interference filters in BSM.

Alice communicates her BSM result to Bob. Bob's unitary operation is composed of two HWPs that change according to Alice's classical information. Finally, after the unitary operation, Bob subjects his photon to polarization dephasing in quartz. Quartz is used in Bob's setup, while YVO$_4$ is used on Alice's side. This is because YVO$_4$ has larger birefringence than quartz, making the thickness of YVO$_4$ thinner than that of quartz with the same decoherence and easing the BSM.

Motor rotating plates and a PBS are used to tomograph the photons and receive the teleported information. The gratings and phase functions both reduce the photon counts. The count of fourfold coincidence detections becomes approximately one tenth, so that the final fourfold coincidence rate is about 0.03 Hz. The data accumulation time for each measurement is $10^4$ s.

**ACKNOWLEDGMENTS**

**Funding:**

The Hefei group was supported by the Innovation Program for Quantum Science and Technology (Grant No. 2021ZD0301200), the National Natural Science Foundation of China (Nos. 62005263, 11821404), the Fundamental Research Funds for the Central Universities (Grant No.WK2030000069), the China Postdoctoral Science Foundation (Grant No. 478 2020M671862). The Turku group acknowledges the financial support from the Magnus Ehrnrooth Foundation and the University of Turku Graduate School (UTUGS).

**Author contributions:**

Z.-D.L. and C.-F.L. planned and designed the experiments. Z.-D.L. implemented the experiments with the help of R.-H.M. and C.-X.N. under the supervision of C.-F.L. and G.-C.G. The theoretical analysis was performed by O.S. under the supervision of T.K. and J.P. The paper was written by Z.-D.L., O.S., T.K., C.-F.L., and J.P. All authors discussed the contents.

**Competing interests:**

Authors declare that they have no competing interests.

**Data and materials availability:**

All data is available in the main text or the Supplementary Materials.

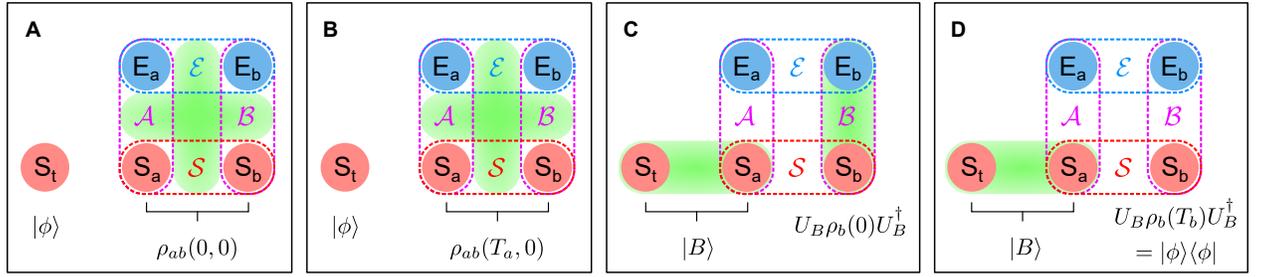

**Fig. 1. Stages of noisy quantum teleportation.** $|\phi\rangle$ is the state to be teleported, initially carried by the system $S_t$. $S_{a(b)}$ is Alice's (Bob's) system/polarization, together forming the composite open system $\mathcal{S}$ that is initially in the state $\rho_{ab}(0,0)$, whose nonlocality remains always hidden, i.e., while there is multipartite hybrid entanglement in the total system, the polarization degree of freedom alone does not display nonlocality. $E_{a(b)}$ is Alice's (Bob's) environment/frequency, together forming the composite environment $\mathcal{E}$. $\mathcal{A}$ ($\mathcal{B}$) is Alice's (Bob's) photon consisting of both the polarization and frequency degrees of freedom, and the green ovals represent entanglement. (**A**) Alice and Bob have shared the auxiliary state $|\Psi(0,0)\rangle$. $\mathcal{A}$ and $\mathcal{B}$ are entangled through the initial probability amplitudes, while $\mathcal{S}$ and $\mathcal{E}$ are entangled through the phase functions $\theta_j(f_j)$. (**B**) Alice subjects her photon to dephasing noise, making the open system evolve from $\rho_{ab}(0,0)$ to $\rho_{ab}(T_a,0)$, whose nonlocality is still hidden. (**C**) Alice performs BSM, entangling $S_t$ and $S_a$. At the same time, the hybrid entanglement transforms into system-environment entanglement on Bob's side, i.e., between $S_b$ and $E_b$. Alice classically communicates her result $|B\rangle$ to Bob, who then, by operating with the matching unitary $U_B$, obtains the polarization state $U_B \rho_b(0) U_B^\dagger$. (**D**) Bob subjects his photon to dephasing noise, which converts the $S_b$–$E_b$ entanglement into coherences within $S_b$, yielding $|\phi\rangle$.

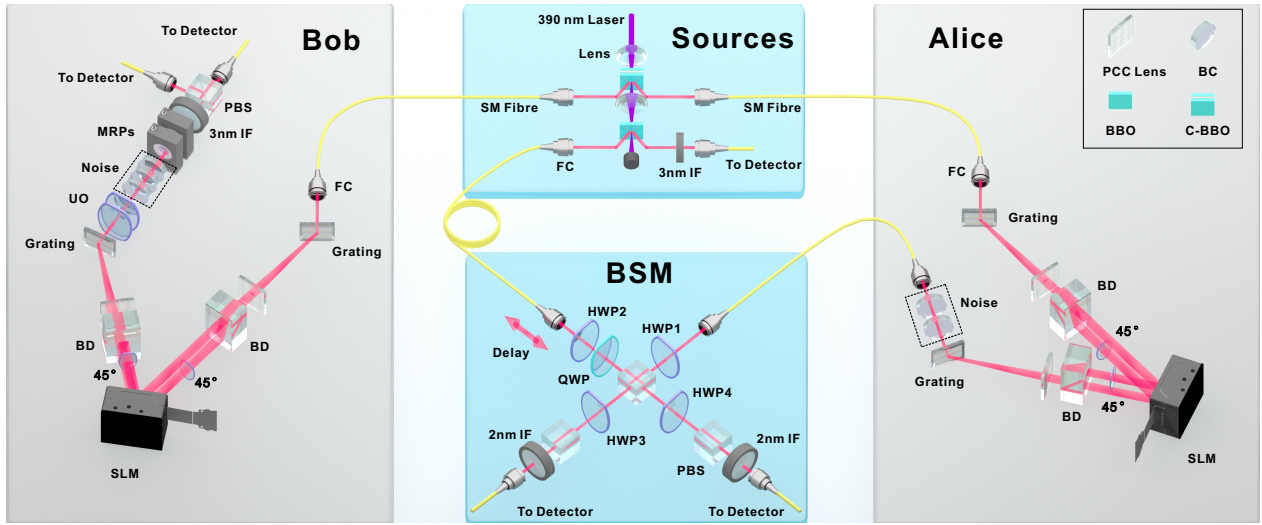

**Fig. 2. Experimental setup.** The setup is composed of four parts. The sources include a polarization entanglement source and a single-photon source. The auxiliary photons' modulation takes place at "Alice" and "Bob". After the modulations, the photons from the single-photon source combine in BSM. Noise is implemented in birefringent crystals (BC). C-BBO—sandwich-like BBO+HWP+BBO combination, BBO—beta barium borate, HWP—half-wave plate, QWP—quarter-wave plate, PCC lens—plano-convex cylindrical lens, BD—beam displacer, MRP—motor rotating plate, PBS—polarizing beam splitter, UO—unitary operation, IF—interference filter, SM fiber—single-mode fiber, FC—fiber collimator.

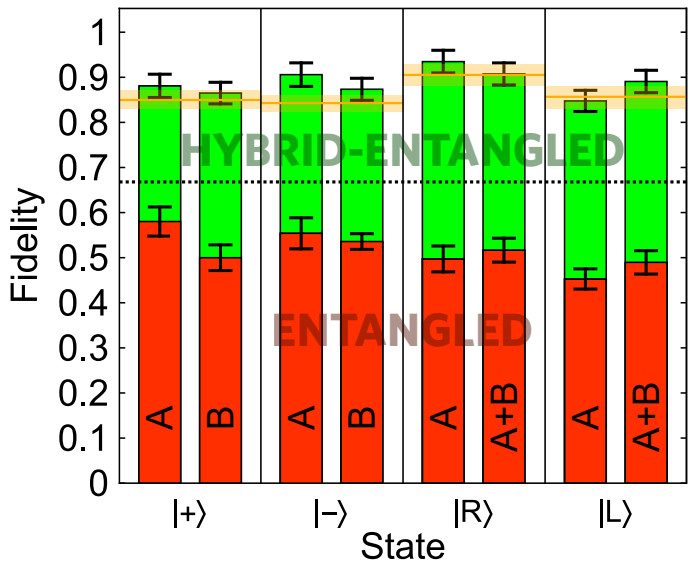

**Fig. 3. Fidelities of the teleported states.** (**A**) Noise only on Alice's side. (**B**) Noise only on Bob's side. (**A+B**) Noise on both sides. The lower, red bars correspond to protocols with standard polarization-entangled initial states. The higher, green bars correspond to protocols with hybrid-entangled initial states. Each panel corresponds to different target

state, which are given below the panels. The orange lines on top of each panel indicate the reference fidelities with no SLMs and no noise. The black dotted line is the classical average fidelity limit (2/3). The error bars are standard deviations calculated by a Monte Carlo method and mainly due to the counting statistics.

# Supplementary Materials for

**Overcoming noise in quantum teleportation with multipartite hybrid entanglement**

Zhao-Di Liu *et al.*

Corresponding authors: Chuan-Feng Li, cfli@ustc.edu.cn; Jyrki Piilo, jyrki.piilo@utu.fi

# Supplementary Materials for

*Overcoming noise in quantum teleportation with multipartite hybrid entanglement*

**Section S1. On hybrid entanglement and hidden nonlocality.**

Here, we elaborate on hybrid entanglement and hidden nonlocality. Let us first show that, following the state preparation, we can go arbitrarily close to zero with the degree of entanglement in polarization, i.e., the degree of freedom being teleported. The bipartite polarization state in question reads

$$\rho_{ab}(t_a, t_b) = \frac{1}{2}\begin{pmatrix} 0 & 0 & 0 & 0 \\ 0 & 1 & \Lambda_{ab}(t_a, t_b) & 0 \\ 0 & \Lambda_{ab}^*(t_a, t_b) & 1 & 0 \\ 0 & 0 & 0 & 0 \end{pmatrix}, \quad (S1)$$

where $\Lambda(t_a, t_b) = \int df_a df_b |g(f_a, f_b)|^2 e^{i[2\pi f_a \Delta n_a (t_a - T_a) - 2\pi f_b \Delta n_b (t_b - T_b)]}$, $t_j$ are the interaction times, and $-T_j$ are the slopes of the linear phase functions, fixed in our experimental setups with spatial light modulators (SLMs).

To quantify entanglement in $\rho_{ab}$, we use the standard definition of concurrence in two-qubit systems, i.e., $C(\rho_{ab}) = \max\{0, \lambda_1 - \lambda_2 - \lambda_3 - \lambda_4\}$, where $\lambda_j$ are decreasingly ordered eigenvalues of the matrix $\sqrt{\sqrt{\rho_{ab}}(\sigma_y \otimes \sigma_y)\rho_{ab}^*(\sigma_y \otimes \sigma_y)\sqrt{\rho_{ab}}}$ (*43*). In our case, one obtains $C(\rho_{ab}(t_a, t_b)) = |\Lambda_{ab}(t_a, t_b)|$. Because we are interested in noise appearing on Bob's side *after* Alice's Bell-state measurement (BSM), we may set $t_b = 0$. By also assuming that $T_j \gg 0$, we get rapidly oscillating phase factors that cancel each other, yielding $|\Lambda_{ab}(t_a, 0)| \approx 0$. Hence, $C(\rho_{ab}(t_a, 0)) \approx 0$ with all interaction times $t_a$, even $t_a = T_a$. If we had $t_j = T_j$ with $j = b$ as well, it would follow from the normalization condition of the frequency spectrum that $|\Lambda_{ab}(T_a, T_b)| = 1$ and thus $C(\rho_{ab}(T_a, T_b)) = 1$. That is, maximum entanglement would emerge from local "filtering", which is the loose definition of hidden nonlocality (*36–39*). However, the nonlocality of the open system remains hidden in our case, meaning that any violation of the Bell-CHSH inequalities (*1*) cannot be experimentally detected during BSM in the bipartite polarization's Hilbert space.

Examining the total system consisting of both the open system and its environment reveals "where" the former's nonlocality is actually hidden. Namely, the total auxiliary system shared by Alice and Bob reads

$$|\Psi(t_a, 0)\rangle = \frac{1}{\sqrt{2}}\{|HV\rangle \int df_a df_b g(f_a, f_b) e^{i[\theta_{aH}(f_a) + 2\pi f_a n_{aH} t_a + \theta_{bV}(f_b)]}|f_a f_b\rangle$$
$$+ |VH\rangle \int df_a df_b g(f_a, f_b) e^{i[\theta_{aV}(f_a) + 2\pi f_a n_{aV} t_a + \theta_{bH}(f_b)]}|f_a f_b\rangle\}. \quad (S2)$$

Clearly $\langle HV|VH\rangle = 0$, but also $\int df_a' df_b' g^*(f_a', f_b') e^{-i[\theta_{aH}(f_a') + 2\pi f_a' n_{aH} t_a + \theta_{bV}(f_b')]} \langle f_a' f_b'|$

$\times \int df_a df_b g(f_a, f_b) e^{i[\theta_{aV}(f_a) + 2\pi f_a n_{aV} t_a + \theta_{bH}(f_b)]} |f_a f_b\rangle = \Lambda_{ab}^*(t_a, 0) \approx 0$ with $T_j \gg 0$.
Therefore, because $|\Psi(t_a, 0)\rangle$ is a balanced superposition of two pairwise orthogonal states, it is fully entangled, and because the states represent different degrees of freedom, one can talk about (multipartite) hybrid entanglement (*33–35*).

Above, we saw that the phase functions determine the degree of hybrid entanglement between the composite open system and the composite environment. Alternatively, one could investigate entanglement between Alice and Bob's subsystems consisting of polarization and frequency. Because both the initial phase functions and the subsequent dephasing can be modeled with local unitaries, the purity of Alice and Bob's subsystems remains constant, and so does the entanglement between them. In this case, concurrence is defined by $C(\Psi_{AB}) = \sqrt{2(1 - \text{tr}[\sigma_A^2])} = \sqrt{2(1 - \text{tr}[\sigma_B^2])}$, where $\sigma_{A(B)}$ is Alice's (Bob's) polarization-frequency state (*44*). Consequently, while the phase functions determine the degree of $\mathcal{S}$–$\mathcal{E}$ entanglement, the initial probability amplitudes determine the degree of $\mathcal{A}$–$\mathcal{B}$ entanglement. Interestingly, it follows that Alice and Bob's photons can be hybrid-entangled without there being any polarization-polarization or frequency-frequency entanglement. It should be mentioned that classical polarization-polarization correlations are still needed for our teleportation protocol to work.

# Supplementary Materials for

*Overcoming noise in quantum teleportation with multipartite hybrid entanglement*

**Section S2. Experimental verification of state purification with "reverse decoherence".**

To test state purification with "reverse decoherence" in a clearer fashion, we purified the entire Bell basis $\{|\Phi^\pm\rangle, |\Psi^\pm\rangle\}$ with different noise configurations. Although these results better illustrate the effect of the phase functions $\boldsymbol{\theta_j(f_j)}$ on the decoherence dynamics, it is important to notice that such purification is *not* needed for our teleportation protocol to work. The experimental setup, which can be obtained from the teleportation setup by simply blocking the pump laser before the BBO, inserting QWP1 on Alice's path, and setting HWP3 and HWP4 to 0°, is shown in Fig. S1.

First, we placed an yttrium orthovanadate (YVO4) plate on Alice's path, matching with the effective path difference $c\Delta n_a T_a = 400\lambda_0$ ($\lambda_0 = 780$ nm). We constructed the phase function $\theta_a(f_a)$ with an SLM preceding the YVO4 plate and changed its slope gradually to cancel the path difference and reach maximum purity. Secondly, we fixed Bob's phase function as $\theta_b(f_b) = -2\pi f_b/c \cdot 429\lambda_0$ and simulated subsequent dephasing by stacking multiple quartz plates on his path.

The resulting fidelities of both configurations are plotted in Fig. S2 with the target Bell states shown in the upper left corners. Note that, in Alice's case, the horizontal axes give the factor $x$ in $\theta_a(f_a) = -2\pi f_a/c \cdot x$ and not the thickness of YVO4. In all cases, the bipartite polarization starts as mixed and ends as pure at approximately $450\lambda_0$ for Alice and $400\lambda_0$ for Bob. The experimental data is in good agreement with the theoretical predictions, which were evaluated numerically by using the actual SLM-pixel values, fitted frequency spectra (see Fig. S3), and Sellmeier equations of the birefringent media.

We suspect that the theory and experiment are in better agreement on Alice's side for two reasons: First, we used only one YVO4 plate in the experiment, which means that small angles between other plates could not affect the results. Secondly, here we imprinted the phase function $\theta_a(f_a)$ on relatively wider frequency domain than on Bob's side, where a wider interference filter (3 nm) was used. That is, more photons *without* the phase function $\theta_b(f_b)$ got through.

Finally, to simulate a more realistic scenario, we fixed the amount of dephasing with a 2 m polarization maintaining single-mode fiber (PM fiber) on Bob's side and purified the state $|\Psi^+\rangle$ by changing the slope of $\theta_b(f_b)$ (see Fig. S4A). We obtained the highest purity with $\theta_b(f_b) = -2\pi f_b/c \cdot 1063\lambda_0$. The amount of noise, $1080\lambda_0$, was used as a fit parameter. The exact amount of noise was not known due to the manufacturer of the fiber not reporting the fiber's Sellmeier equation. Hence, dispersion was not accounted for, which explains the difference between the maxima of theory and experiment. To further visualize our protocol, we have plotted the real values of the initial and time-evolved density matrix elements in Figs. S4B and S4C, respectively. The imaginary values were negligibly small.

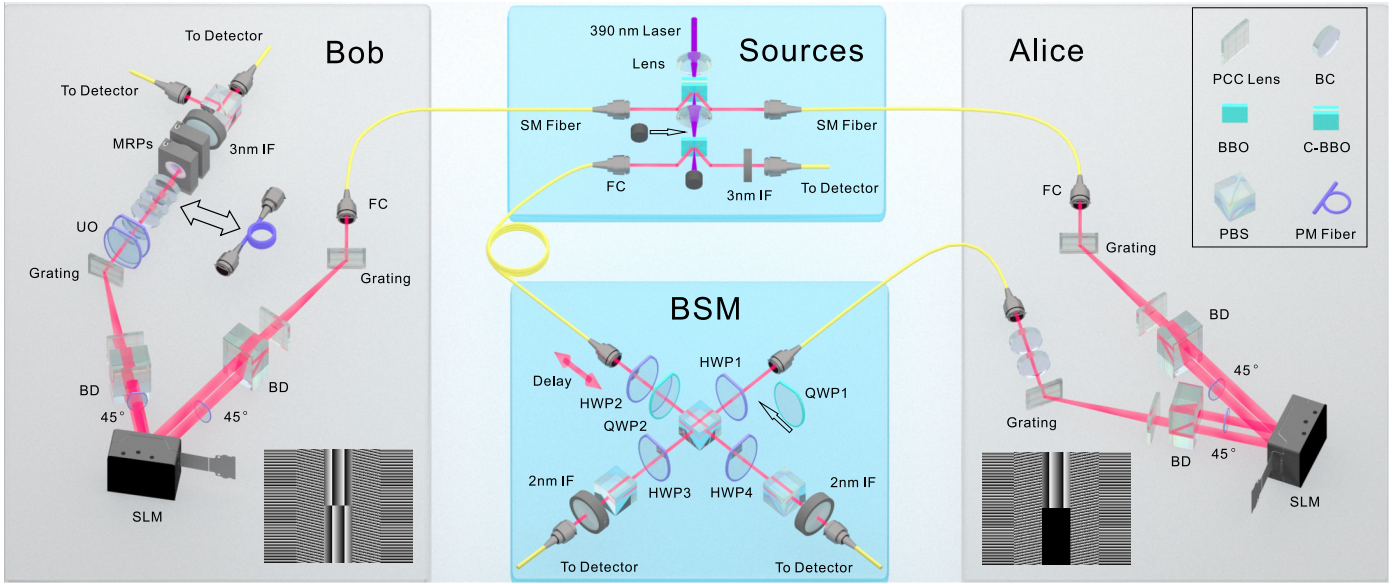

**Fig. S1. Experimental setup of state purification.** The setup can be changed from noisy quantum teleportation to state purification by blocking the 390 nm pump laser between C-BBO (sandwich-like BBO+HWP+BBO combination) and BBO (beta barium borate) and inserting QWP1 (quarter-wave plate) in the setup. Here, we also use a PM fiber (polarization maintaining single-mode fiber) on Bob's side, corresponding to "real-life noise". Alice and Bob's SLMs are accompanied by sample holograms in the picture. The sample hologram on Alice's side matches with $400\lambda_0$ of YVO$_4$, while the sample hologram on Bob's side matches with 2 m polarization maintaining single-mode fiber (PM fiber). HWP—half-wave plate, PCC lens—plano-convex cylindrical lens, BD—beam displacer, BC—birefringent crystal, MRP—motor rotating plate, PBS— polarizing beam splitter, UO—unitary operation, IF—interference filter, SM fiber—single-mode fiber, FC—fiber collimator.

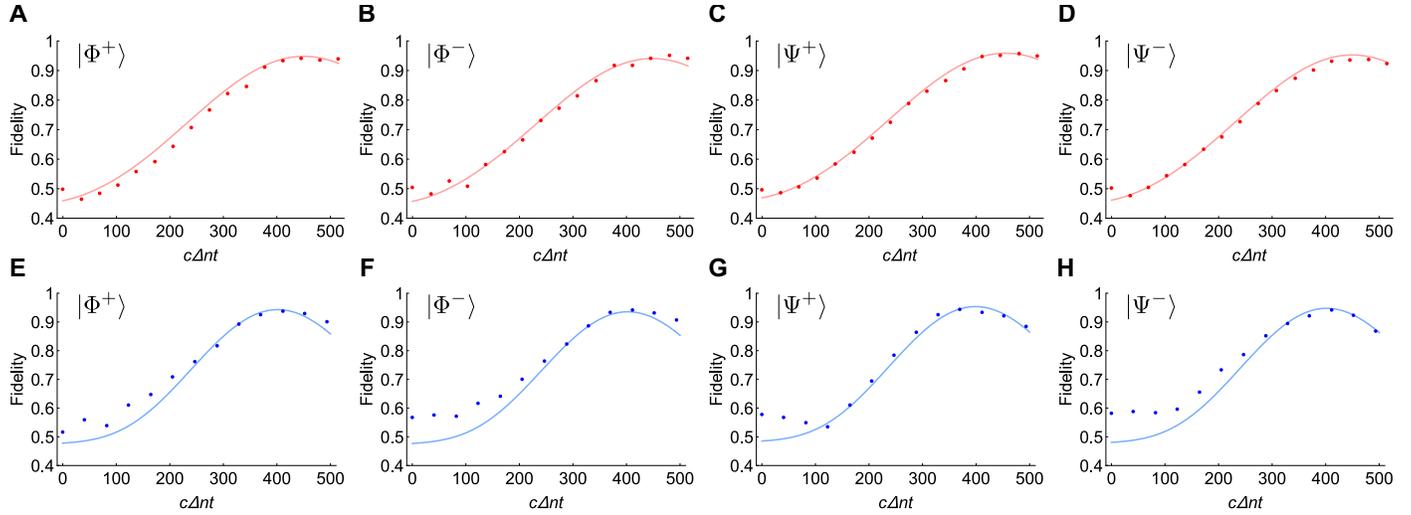

**Fig. S2. Results of state purification with "reverse decoherence".** Fidelities of the purified states as functions of the effective path difference in units of $\boldsymbol{\lambda_0}$. The target Bell states are shown in the upper left corners. The solid curves represent the theoretical predictions, while the dots correspond to the measurement data. The error bars are standard deviations calculated by a Monte Carlo method and of the same size as the dots. The numerical values of the largest error bars in each of the panels are: (**A**) 0.0040, (**B**) 0.0039, (**C**) 0.0055, (**D**) 0.0040, (**E**) 0.0048, (**F**) 0.0044, (**G**) 0.0035, and (**H**) 0.0046.

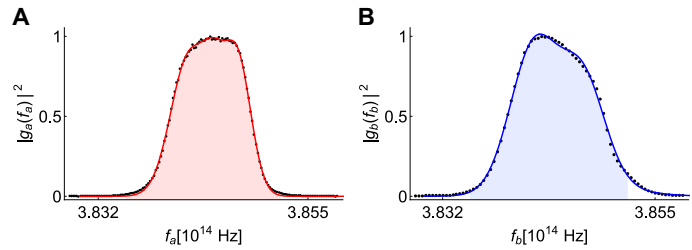

**Fig. S3. The frequency spectra.** (**A**) Alice's margin of the joint frequency spectrum after the 2 nm interference filter. (**B**) Bob's margin of the joint frequency spectrum after the 3 nm interference filter. The black dots are the measured (and scaled) counts. The solid curves are the fitted functions, i.e., the weighted sums of three Gaussians. The filled areas represent the frequency domains of the phase functions.

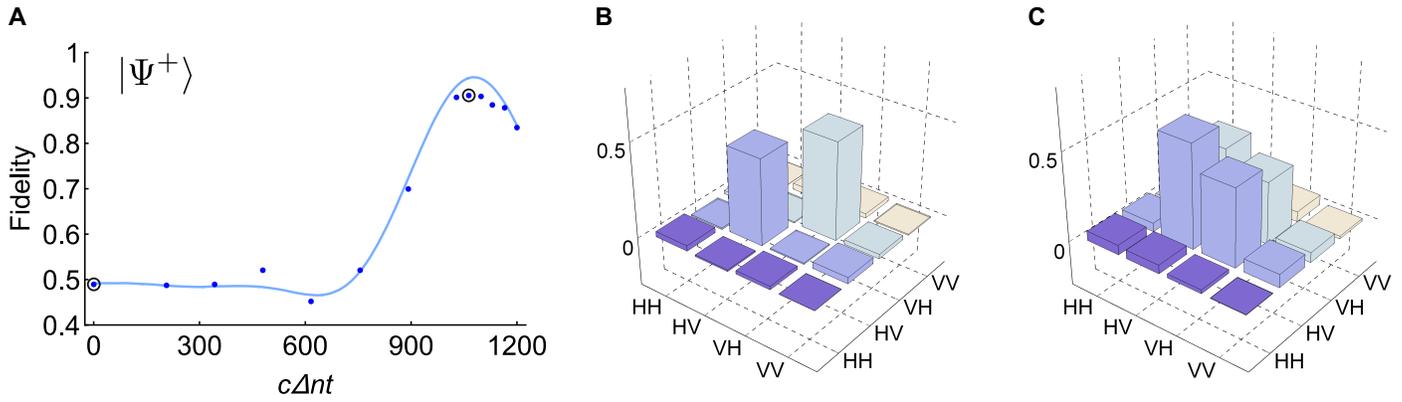

**Fig. S4. State purification in an optical fiber. (A)** Fidelity of the purified state $|\Psi^+\rangle$ as a function of the effective path difference in units of $\lambda_0$. The solid curve represents the theoretical prediction, while the dots correspond to the measurement data. The error bars are standard deviations calculated by a Monte Carlo method and of the same size or smaller as the dots. The numerical value of the largest error bar is 0.0049. **(B)**, **(C)** Real values of **(B)** the initial and **(C)** time-evolved density matrix elements, corresponding to the circled data points in panel **(A)**.